# Current-nonlinear Hall effect and spin-orbit torque magnetization switching in a magnetic topological insulator

K. Yasuda[1*], A. Tsukazaki[2], R. Yoshimi[3], K. Kondou[3],

K. S. Takahashi[3], Y. Otani[3,4], M. Kawasaki[1,3], and Y. Tokura[1,3]

[1] *Department of Applied Physics and Quantum-Phase Electronics Center (QPEC),*

*University of Tokyo, Tokyo 113-8656, Japan*

[2] *Institute for Materials Research, Tohoku University, Sendai 980-8577, Japan*

[3] *RIKEN Center for Emergent Matter Science (CEMS), Wako 351-0198, Japan*

[4] *Institute for Solid State Physics (ISSP), University of Tokyo, Kashiwa 277-8581, Japan.*

* Corresponding author: yasuda@cmr.t.u-tokyo.ac.jp

# ABSTRACT

**Precise estimation of spin Hall angle as well as successful maximization of spin-orbit torque (SOT) form a basis of electronic control of magnetic properties with spintronic functionality. Until now, current-nonlinear Hall effect, or second harmonic Hall voltage has been utilized as one of the methods for estimating spin Hall angle, which is attributed to the magnetization oscillation by SOT. Here, we argue the second harmonic Hall voltage in magnetic/nonmagnetic topological insulator (TI) heterostructures, $Cr_x(Bi_{1-y}Sb_y)_{2-x}Te_3/(Bi_{1-y}Sb_y)_2Te_3$. From the angular, temperature and magnetic field dependence, it is unambiguously shown that the large second harmonic Hall voltage in TI heterostructures is governed not by SOT but mainly by asymmetric magnon scattering mechanism without magnetization oscillation. Thus, this method does not allow an accurate estimation of spin Hall angle when magnons largely contribute to electron scattering. Instead, the SOT contribution in a TI heterostructure is exemplified by current pulse induced non-volatile magnetization switching, which is realized with a current density of ~ $2.5 \times 10^{10}$ A/m$^2$, showing its potential as spintronic materials.**

# MAIN TEXT

## INTRODUCTION

Interconversion of angular momentum between conduction electron and local magnetization is one of the central issues of contemporary spintronic research. For example, in normal metal/ferromagnet (NM/FM) heterostructures, accumulated spins via spin current in NM play a key role in manipulating magnetization of FM, which is referred to as spin-orbit torque (SOT) (*1*, *2*). This enables magnetization switching (*3-6*) as well as fast domain wall motion (*7*, *8*), which directly leads to computation, logic and memory device applications. In particular, materials with large spin Hall angle $\theta_{SH}$ can provide large spin current with minimal Joule heating enabling energy-saving spintronic devices in the future. Reflecting the importance of an accurate estimation of $\theta_{SH}$, several techniques have recently been developed such as spin pumping (*9*, *10*), spin torque ferromagnetic resonance (ST-FMR) (*4*, *5*, *11-14*) and second harmonic Hall voltage measurement (*15-19*). Utilizing these methods, three-dimensional topological insulators (TI) have been demonstrated to have quite a large $\theta_{SH}$ (*10*, *12-14*, *18*, *19*), a few times to two orders of magnitude larger than heavy metal elements such as Pt (*4*), $\beta$-Ta (*5*) and $\beta$-W (*6*).

TI is a class of materials with insulating bulk and conductive surface state with Dirac dispersion (*20*). The spin direction of the surface electron is locked to its momentum, termed "spin-momentum locking". This one-to-one correspondence between charge and spin degrees of freedom makes TI an ideal spintronic material with large $\theta_{SH}$ (*10*, *12-14*, *18*, *19*). Moreover, magnetic TI provides one of the ideal platforms to exemplify the spin-polarized electron transport interacting with underlying magnetism (*18*, *19*, *21-25*). In fact, large current-direction-dependent or unidirectional magnetoresistance (UMR) (*26-29*) is observed under appropriate magnetization and current directions (*25*). Such current-nonlinear longitudinal resistance is attributed to the asymmetric scattering of conduction electron by magnon due to the conservation of angular momentum, what we call "asymmetric magnon scattering mechanism" (*25*).

In this paper, we reveal that the asymmetric magnon scattering mechanism gives rise to current-nonlinear resistance not only in the longitudinal direction but also in the transverse direction in magnetic TI under certain configuration, which is observed as the large second harmonic Hall voltage. Importantly, the present result means that $\theta_{SH}$ cannot be accurately evaluated by second harmonic technique because the second harmonic voltage is governed not by the SOT but mainly by the asymmetric magnon scattering. As a target material, we characterize TI heterostructures (*18*, *19*, *24*, *25*, *30*) composed of nonmagnetic TI $(Bi_{1-y}Sb_y)_2Te_3$ (BST) (*31*, *32*) and magnetic TI $Cr_x(Bi_{1-y}Sb_y)_{2-x}Te_3$ (CBST) (*21-23*) on semi-insulating InP substrate. By changing the Bi to Sb ratio (composition *y*), we tune the Fermi energy $E_F$ of the surface state with single Dirac cone at the Γ point (*20*). Hence, only top and bottom surfaces are conductive as can be seen from the observation of quantum Hall effect under high perpendicular magnetic field (see section S1) (*30*, *32*). Here, the second harmonic voltage is expected only from one surface involved in the Cr-doped magnetic layer, which would be otherwise canceled out by the opposite contributions of top and bottom surfaces (*25*, *30*).

Anomalous Hall voltage is usually proportional to current *J* and out-of-plane component of magnetization $M_z$. In addition, at $J \parallel M \parallel x$ configuration in the heterostructure (Fig. 1B), an additional nonlinear transverse voltage proportional to $J^2$ is allowed from symmetry. Here, the transverse voltage is expressed as,

$$V_y = R_{AHE}J_xM_z + R_{yx}^{(2)}J_x^2, \qquad (1)$$

where $R_{yx}^{(2)}$ is a coefficient of the $J^2$-proportional term. Note that we can neglect ordinary Hall effect and planar Hall effect since they are prohibited in the present experimental configuration and hence irrelevant to the present discussion. The first term of equation (1), corresponding to anomalous Hall voltage, gets zero when *M* is pointing along the in-plane direction (*x*). When a large current is applied, however, *M* is tilted to the out-of-plane direction by anti-damping torque (*1*, *2*) as shown in Fig. 1A so that $M_z = c_{SOT}J_x$, where $c_{SOT}$ is a proportional constant. This is because the effective field by SOT ($B_{SOT}$) is directed along $\sigma \times M$ (3, 4), where $\sigma$ is the spin accumulation direction due to the Rashba-Edelstein effect (33). Therefore, under large current and $J \parallel B \parallel x$ configuration, the Hall voltage is expressed as,

$$V_y = R_{AHE} c_{SOT} J_x^2 + R_{yx}^{(2)} J_x^2. \qquad (2)$$

Both of these two terms are proportional to $J^2$ (current-nonlinear) but the physical meaning is completely different; $M$ is tilted by the SOT mechanism whereas $M$ is unaffected in nonlinear conduction. Only when the second term in equation (2) is negligible, $V_y \simeq R_{AHE} c_{SOT} J_x^2$ allows us to estimate $B_{SOT}$ through the second harmonic Hall voltage (15-19) or current-direction-dependent Hall resistance (4, 5).

## RESULTS

### Observation of large nonlinear Hall voltage

In a second harmonic Hall voltage measurement, we put ac current $J^\omega = J\sin(\omega t)$ in the $x$ direction, then the transverse voltage $V_y = (R_{AHE} c_{SOT} + R_{yx}^{(2)})(\frac{J^2}{2} - \frac{J^2}{2}\cos(2\omega t))$ will appear, which can be measured by lock-in technique. Figure 1C shows the magnetic field dependence of second harmonic Hall voltage $V_y^{2\omega}$ for the CBST(3 nm)/BST(5 nm)/InP film. At $B = 0$ T, magnetization of the film points along $z$ direction due to the perpendicular anisotropy of the CBST/BST heterostructure (18). Here, $V_y^{2\omega}$ is almost zero with small hysteresis. As magnetic field is applied up to ~ 0.7 T, $M$ points along $x$ direction. After taking the maximum at ~ 0.7 T, $V_y^{2\omega}$ decreases with increasing magnetic field. Moreover, as shown in Fig. 1D, the sign of $V_y^{2\omega}$ is reversed in the inverted heterostructure, BST(3 nm)/CBST(5 nm)/InP, while showing a comparable magnitude. This is because the manner of the spin-momentum locking is opposite between the top and bottom surface states. Figure 1E shows the current amplitude dependence of $V_y^{2\omega}$. At low current amplitude, $V_y^{2\omega}$ is proportional to $J^2$ as expected from equation (2). The deviation from the proportionality at larger current originates from the heating effect (25). When we assume that $V_y^{2\omega}$ originates only from the SOT mechanism, namely from the first term in equation (2), the effective field $B_{SOT}$ would amount to ~ 26 mT under $J = 1$ μA. As a consequence, $\theta_{SH}$ would be estimated to be ~ 160, assuming the conductive regions of top and bottom surfaces to be 1 nm (*34*) (see section S2 for the estimation of $B_{SOT}$ and $\theta_{SH}$). The estimated $\theta_{SH}$ is comparable to that deduced by the previous second harmonic measurements ($\theta_{SH}$ : 80 - 425) (*18*, *19*) and extraordinarily larger than estimations by ST-FMR or spin pumping ($\theta_{SH}$ : 0.4 - 3.5) (*10*, *12-14*). As shown in the following, we argue against the extraordinary value of $\theta_{SH}$ as derived from the observed large $V_y^{2\omega}$.

## Origin of nonlinear Hall voltage

To examine the origin of large $V_y^{2\omega}$ signal, we measured angular, temperature and magnetic field dependence of $V_x^{2\omega}$ and $V_y^{2\omega}$ in CBST/BST/InP. Figure 2, (A to F) show the angular dependence of second harmonic longitudinal voltage $V_x^{2\omega}$ and second harmonic Hall (transverse) voltage $V_y^{2\omega}$. As can be seen from the detailed angular dependence in $zx$ plane (Fig. 2, A and D), $zy$ plane (Fig. 2, B and E) and $xy$ plane (Fig. 2, C and F), $V_y^{2\omega}$ is largest when $M \parallel J \parallel x$. First, let us assume that $V_y^{2\omega}$ originated purely from SOT, namely from magnetization oscillation. In this case, $V_x^{2\omega}$ in Fig. 2D should become non-zero as well, because longitudinal resistance $R_{xx}$ depends on the magnetization direction; $R_{xx}$ amounts to 16.4 kΩ (15.6 kΩ) when $M \parallel z$ ($M \parallel x$). When $M$ is oscillating in the $zx$ plane, it would modulate longitudinal resistance as well as anomalous Hall resistance so that $V_x^{2\omega}$ should also get finite (*17*). Here, the calculated signal from SOT $V_x^{2\omega, SOT}$ is shown in green line in Fig. 2D (the detailed derivation is shown in section S3). The experimental results, however, does not show such a signal within an experimental error. Therefore, we can conclude that SOT least contributes to $V_x^{2\omega}$ and eventually to $V_y^{2\omega}$, namely $V_y^{2\omega}$ should originate mainly from the $J^2$-proportional nonlinear conduction (the second term in equation. (2)).

To make further discussions, the microscopic origin of $J^2$-proportional conduction has to be revealed. In this context, we notice that $V_y^{2\omega}$ and $V_x^{2\omega}$ have duality; the magnitude of the signal and the angular dependence have a correspondence. Therefore, it is natural to conjecture that $V_x^{2\omega}$ and $V_y^{2\omega}$ possess the same microscopic origin. In Fig. 2, (D to F), $V_x^{2\omega}$ gets maximum when $M \perp J$. This is nothing but UMR (25-29), which comes from the second term of $V_x = R_{xx}J_x + R_{xx}^{(2)}J_x^2$ (25). Here, the $J^2$-proportional voltage derives from the asymmetric relaxation time between electron with positive group velocity (position A) and that with negative group velocity (C) as shown in Fig. 2G (25): The scattering processes by magnon are inequivalent between A and C due to spin-momentum locking and conservation of angular momentum (25, 29). In a similar manner, the $J^2$-proportional transverse voltage ($R_{yx}^{(2)}J_x^2$ in equations (1) and (2)) can be derived when $M \parallel J$ as shown in Fig. 2H. When electron with positive group velocity (at around position B, angular momentum +1/2 with the quantization direction is taken along $x \parallel M$) is scattered to around D (angular momentum −1/2), the process accompanies the emission of magnon with the angular momentum of +1. On the other hand, the scattering of electron with negative group velocity (at around D, −1/2) to around B (+1/2) must absorb magnon. This nonequivalence in scattering processes leads to the asymmetry in relaxation time, which is the origin of the $J^2$-proportional transverse term (nonlinear conduction), as schematically shown in Fig. 1B. Such asymmetric magnon scattering model allows us to evaluate $R_{yx}^{(2)}$; its detail derivation is shown in section S4. From that model, we can analytically show that $V_x^{2\omega}(J \parallel x, M \parallel y) = -V_y^{2\omega}(J \parallel x, M \parallel x)/3$; namely $V_x^{2\omega}$ and $V_y^{2\omega}$ show the same order of magnitude, consistent with the experimental results including their signs (see section S5). The quantitative deviation of the ratio $V_x^{2\omega}/V_y^{2\omega}$ between the calculation and the experiment originates from the approximations adopted in the calculation such as relaxation time approximation, simplified cone band of the surface state and dispersionless magnon band (25).

To further examine the validity of the asymmetric magnon scattering model, we calculate the temperature and magnetic field dependence of $V_y^{2\omega}$ and compare with experimental results in Fig. 3. The experiment and calculation show qualitative consistency in a similar way to those observed in UMR (or $V_x^{2\omega}$) (25); $V_y^{2\omega}$ monotonically decreases as raising temperature at low magnetic fields ($B$ = 1 T and 5 T) and almost vanishes at around $T_c$ ~ 34 K. On the other hand, $V_y^{2\omega}$ at a high magnetic field (13.9 T) takes a peak structure at around 5 K both in experiment and calculation. According to the asymmetric magnon scattering model, such a change in the typical energy scale as a function of magnetic field $B$ can be understood as the variation of the magnon gap (~ $g\mu_B B$, corresponding to 20 K at 13.9 T), where $g$ and $\mu_B$ are $g$ factor ($g \simeq 2$ for $Cr^{3+}$) and Bohr magneton, respectively. In contrast, such a change of the energy scale cannot be explained with the SOT-based model, since SOT has no characteristic energy scale (2). Therefore, the results again support the asymmetric magnon scattering origin of $V_y^{2\omega}$, namely $R_{AHE} c_{SOT} \ll R_{yx}^{(2)}$, which leads to the overestimation of $\theta_{SH}$ in magnetic TI.

**Current-pulse induced non-volatile magnetization switching by SOT**

Then, in what current region and how can the effect of SOT show up for the present TI heterostructure? Note here that the above argument of nonlinear Hall effect is applicable not only to the ac current measurement but also to the dc current one. Since $R_{AHE} c_{SOT} \ll R_{yx}^{(2)}$, the Hall resistance measured under dc current and large in-plane magnetic field becomes $R_{yx} = V_y / J_x \cong R_{yx}^{(2)} J_x$. Such current-direction-dependent Hall resistance due to the asymmetric magnon scattering is hard to distinguish from SOT contribution (18, 19). To avoid this difficulty, the non-volatile magnetization switching has to be examined directly by current pulse injection, but not by Hall resistance measurement under large dc current. Figure 4A shows the schematic illustration of current induced magnetization switching experiment; current pulse $J_{pulse}$ is injected parallel to the in-plane external magnetic field $B$. Here, the effective field $B_{SOT}$ rotates the magnetization with the assistance of small in-plane $B$, which is required to define the switching direction. The current pulse injection scheme is shown in Fig. 4B. After every current pulse injection (gray bar), the possible change in $M_z$ is evaluated though Hall resistance $R_{yx}$ measured with low current (1 µA, green triangle) (*3*). Here, the asymmetric magnon scattering is negligible because its contribution of ~ 10 µV is much smaller than the anomalous Hall voltage of ~ 1 mV. Figure 4C shows $R_{yx}$ - $J_{pulse}$ loops at 2 K under in-plane magnetic field $B = \pm 0.02$ T. Note that 0.02 T is much smaller than anisotropy field 0.7 T and therefore magnetization is pointing almost exclusively up or down. Thus, the intermediate value of $R_{yx}$ or normalized $M_z$ indicates the multidomain formation. As clearly seen in Fig. 4C, $R_{yx}$ is switched from negative (positive) to positive (negative) at around $J_{pulse}$ = –0.5 mA (+0.5 mA) under $B$ = 0.02 T, pointing to the current induced magnetization reversal. The threshold current $J_{th}$ = 0.5 mA corresponds to current density of $2.5 \times 10^{10}$ A/m$^2$. The switching direction is reversed under $B$ = –0.02 T. Furthermore, the magnetization direction can be repeatedly controlled by current injection of ±1 mA as shown in Fig. 4D. We note that the magnetization switching ratio (normalized $M_z$) is limited to ~ 0.4, which is relatively small compared with the full switching operation in NM/FM heterostructures (*3-6*), whose origin is yet to be clarified. At present, we speculate that some domains are hard to switch by current pulse due to the strong pinning.

## DISCUSSION

The large nonlinear Hall effect in magnetic/nonmagnetic TI heterostructures is shown to be governed not by SOT but mainly by asymmetric magnon scattering. Therefore, if the spin Hall angle $\theta_{SH}$ in magnetic TI heterostructure were deduced from the second harmonic Hall voltage measurement without careful consideration of asymmetric magnon scattering (*18, 19*), it would lead to the overestimation of $\theta_{SH}$ value. In this analogy, the asymmetric magnon scattering contribution to the second harmonic Hall voltage might not be negligible in conventional NM/FM heterostructures (*15-17*) such as Pt/Py, because the UMR based on the asymmetric magnon scattering mechanism is observed in it as well (*29*).

The true SOT contribution in magnetic TI heterostructure is studied by the current-pulse induced non-volatile magnetization switching, which is realized at a current density of ~ $2.5 \times 10^{10}$ A/m$^2$. In spite of the larger perpendicular magnetic anisotropy, the threshold current density is still much smaller than those ($10^{11}$ - $10^{12}$ A/m$^2$) of NM/FM heterostructures (*3-6*). This can be attributed to the smaller saturation magnetization $M_s = 3.5 \times 10^4$ A/m as well as to the large $\theta_{SH}$ of TI (*10*, *12-14*) (of orders of one, see also section S6), showing its high potential as spintronic materials.

# MATERIALS AND METHODS

## Thin film growth and device fabrication

The TI heterostructures were grown on semi-insulating InP substrates with molecular beam epitaxy (MBE) in the same procedures as described in Refs. *24* and *30*. The nominal compositions of TI heterostructure Cr$_x$(Bi$_{1-y}$Sb$_y$)$_{2-x}$Te$_3$/(Bi$_{1-y}$Sb$_y$)$_2$Te$_3$ are $x$ ~ 0.2 and $y$ ~ 0.88. Using photolithography and Ar ion milling or chemical etching, the heterostructures were patterned into the shape of Hall bars with 10 μm in width and 20 μm in length. The electrode Au (45 nm)/Ti (5 nm) was formed by electron beam deposition.

## Second harmonic voltage measurements

The second harmonic voltages were measured using a current source (Keithley: Model 6221) and lock-in amplifiers (SRS: SR830). The measurement frequency was fixed to 13 Hz. Most of the measurements were done at 2 K in Physical Property Measurement System (Quantum Design: PPMS) unless otherwise noted. The second harmonic voltages are anti-symmetrized as a function of *B* and *M*.

## Current induced magnetization switching

The pulse current injection with varying pulse heights and the subsequent Hall resistance measurement with enough low current (~ 1 μA) were done with a current source (Keithley: Model 6221) and a voltmeter (Keithley: Model 2182A). The pulse width was set to be ~ 1 ms.

# SUPPLEMENTARY MATERIALS

Section S1. Transport properties of the sample.

Section S2. Estimation of spin Hall angle with the assumption that $V_x^{2\omega}$ is purely derived from SOT.

Section S3. Simulation of $V_x^{2\omega,\text{SOT}}$ with SOT origin.

Section S4. Derivation of $R_{yx}^{(2)}$.

Section S5. Relationship between $V_x^{2\omega}$ and $V_y^{2\omega}$.

Section S6. Estimation of spin Hall angle by current-pulse-induced magnetization switching.

Fig. S1. Transport properties of the sample.

Fig. S2. Fitting of $V_x^{2\omega}$ with the assumption that it is purely derived from SOT.

Fig. S3. In-plane magnetic field dependence of $R_{xx}$.

Fig. S4. Schematic top view of the magnon scattering process in 2D Dirac dispersion.

# REFERENCES AND NOTES

1. J. C. Slonczewski, Current-driven excitation of magnetic multilayers. *J. Magn. Magn. Mat.* **159**, L1-L7 (1996).
2. D. C. Ralph, M. D. Stiles, Spin transfer torques. *J. Mag. Mag. Mat.* **320**, 1190-1216 (2008).
3. I. M. Miron, K. Garello, G. Gaudin, P. J. Zermatten, M. V. Costache, S. Auffret, S. Bandiera, B. Rodmacq, A. Schuhl, P. Gambardella, Perpendicular switching of a single ferromagnetic layer induced by in-plane current injection. *Nature* **476**, 189–193 (2011).
4. L. Liu, O. J. Lee, T. J. Gudmundsen, D. C. Ralph, R. A. Buhrman, Current-induced switching of

perpendicularly magnetized magnetic layers using spin torque from the spin Hall effect. *Phys. Rev. Lett.* **109**, 096602 (2012).

5. L. Liu, C. F. Pai, Y. Li, H. W. Tseng, D. C. Ralph, R. A. Buhrman, Spin-torque switching with the giant spin Hall effect of Tantalum. *Science* **336**, 555-558 (2012).

6. C. F. Pai, L. Liu, Y. Li, H. W. Tseng, D. C. Ralph, R. A. Buhrman, Spin transfer torque devices utilizing the giant spin Hall effect of tungsten. *Appl. Phys. Lett.* **101**, 122404 (2012).

7. S. Emori, U. Bauer, S. M. Ahn, E. Martinez, G. S. D. Beach, Current-driven dynamics of chiral ferromagnetic domain walls. *Nat. Mater.* **13**, 611–616 (2013).

8. K. S. Ryu, L. Thomas, S. H. Yang, S. Parkin, Chiral spin torque at magnetic domain walls. *Nat. Nanotech.* **8**, 527–533 (2013).

9. O. Mosendz, J. E. Pearson, F. Y. Fradin, G. E. W. Bauer, S. D. Bader, A. Hoffmann, Quantifying spin Hall angles from spin pumping: Experiments and theory. *Phys. Rev. Lett.* **104**, 046601 (2010).

10. M. Jamali, J. S. Lee, J. S. Jeong, F. Mahfouzi, Y. Lv, Z. Zhao, B. K. Nikolić, K. A. Mkhoyan, N. Samarth, J. P. Wang, Giant spin pumping and inverse spin Hall effect in the presence of surface and bulk spin−orbit coupling of topological insulator $Bi_2Se_3$. *Nano Lett.* **15**, 7126−7132 (2015).

11. L. Liu, T. Moriyama, D. C. Ralph, R. A. Buhrman, Spin-torque ferromagnetic resonance induced by the spin Hall effect. *Phys. Rev. Lett.* **106**, 036601 (2011).

12. A. R. Mellnik, J. S. Lee, A. Richardella, J. L. Grab, P. J. Mintun, M. H. Fischer, A. Vaezi, A. Manchon, E. A. Kim, N. Samarth, D. C. Ralph, Spin-transfer torque generated by a topological insulator. *Nature* **511**, 449–451 (2014).

13. Y. Wang, P. Deorani, K. Banerjee, N. Koirala, M. Brahlek, S. Oh, H. Yang, Topological surface states originated spin-orbit torques in $Bi_2Se_3$. *Phys. Rev. Lett.* **114**, 257202 (2015).

14. K. Kondou, R. Yoshimi, A. Tsukazaki, Y. Fukuma, J. Matsuno, K. S. Takahashi, M. Kawasaki, Y. Tokura, Y. Otani, Fermi-level-dependent charge-to-spin current conversion by Dirac surface states of topological insulators. *Nat. Phys.* **12**, 1027–1031 (2016).

15. J. Kim, J. Sinha, M. Hayashi, M. Yamanouchi, S. Fukami, T. Suzuki, S. Mitani, H. Ohno, Layer thickness dependence of the current-induced effective field vector in Ta|CoFeB|MgO. *Nat. Mater.* **12**, 240-245 (2013).

16. K. Garello, I. M. Miron, C. O. Avci, F. Freimuth, Y. Mokrousov, S. Blügel, S. Auffret, O. Boulle, G. Gaudin, P. Gambardella, Symmetry and magnitude of spin–orbit torques in ferromagnetic heterostructures. *Nat. Nanotech.* **8**, 587-593 (2013).

17. M. Hayashi, J. Kim, M. Yamanouchi, H. Ohno, Quantitative characterization of the spin-orbit torque using harmonic Hall voltage measurements. *Phys. Rev. B* **89**, 144425 (2014).

18. Y. Fan, P. Upadhyaya, X. Kou, M. Lang, S. Takei, Z. Wang, J. Tang, L. He, L. T. Chang, M. Montazeri, G. Yu, W. Jiang, T. Nie, R. N. Schwartz, Y. Tserkovnyak, K. L. Wang, Magnetization switching through giant spin–orbit torque in a magnetically doped topological insulator heterostructure. *Nat. Mater.* **13**, 699–704 (2014).

19. Y. Fan, X. Kou, P. Upadhyaya, Q. Shao, L. Pan, M. Lang, X. Che, J. Tang, M. Montazeri, K. Murata, L. T. Chang, M. Akyol, G. Yu, T. Nie, K. L. Wong, J. Liu, Y. Wang, Y. Tserkovnyak, K. L. Wang, Electric-field control of spin–orbit torque in a magnetically doped topological insulator. *Nat. Nanotech.* **11**, 352–359 (2016).

20. M. Z. Hasan, C. L. Kane, Colloquium: Topological insulators. *Rev. Mod. Phys.* **82**, 3045 (2010).

21. C. Z. Chang, J. Zhang, X. Feng, J. Shen, Z. Zhang, M. Guo, K. Li, Y. Ou, P. Wei, L. L. Wang, Z. Q. Ji, Y. Feng, S. Ji, X. Chen, J. Jia, X. Dai, Z. Fang, S. C. Zhang, K. He, Y. Wang, L. Lu, X. C. Ma, Q. K. Xue, Experimental observation of the quantum anomalous Hall effect in a magnetic topological insulator. *Science* **340**, 167-170 (2013).

22. J. G. Checkelsky, R. Yoshimi, A. Tsukazaki, K. S. Takahashi, Y. Kozuka, J. Falson, M. Kawasaki, Y. Tokura, Trajectory of the anomalous Hall effect towards the quantized state in a ferromagnetic topological insulator. *Nat. Phys.* **10**, 731–736 (2014).

23. X. Kou, S. T. Guo, Y. Fan, L. Pan, M. Lang, Y. Jiang, Q. Shao, T. Nie, K. Murata, J. Tang, Y.

Wang, L. He, T. K. Lee, W. L. Lee, K. L. Wang, Scale-Invariant Quantum Anomalous Hall Effect in Magnetic Topological Insulators beyond the Two-Dimensional Limit. *Phys. Rev. Lett.* **113**, 137201 (2014).

24. K. Yasuda, R. Wakatsuki, T. Morimoto, R. Yoshimi, A. Tsukazaki, K. S. Takahashi, M. Ezawa, M. Kawasaki, N. Nagaosa, Y. Tokura, Geometric Hall effects in topological insulator heterostructures. *Nat. Phys.* **12**, 555-559 (2016).

25. K. Yasuda, A. Tsukazaki, R. Yoshimi, K. S. Takahashi, M. Kawasaki, Y. Tokura, Large unidirectional magnetoresistance in a magnetic topological insulator. *Phys. Rev. Lett.* **117**, 127202 (2016).

26. K. Olejník, V. Novák, J. Wunderlich, T. Jungwirth, Electrical detection of magnetization reversal without auxiliary magnets. *Phys. Rev. B* **91**, 180402 (2015).

27. C. O. Avci, K. Garello, A. Ghosh, M. Gabureac, S. F. Alvarado, P. Gambardella, Unidirectional spin Hall magnetoresistance in ferromagnet/normal metal bilayers. *Nat. Phys.* **11**, 570-575 (2015).

28. C. O. Avci, K. Garello, J. Mendil, A. Ghosh, N. Blasakis, M. Gabureac, M. Trassin, M. Fiebig, P. Gambardella, Magnetoresistance of heavy and light metal/ferromagnet bilayers. *Appl. Phys. Lett.* **107**, 192405 (2015).

29. K. J. Kim, T. Moriyama, T. Koyama, D. Chiba, S. W. Lee, S. J. Lee, K. J. Lee, H. W. Lee, T. Ono Current-induced asymmetric magnetoresistance due to energy transfer via quantum spin-flip process. Preprint at https://arxiv.org/abs/1603.08746 (2016).

30. R. Yoshimi, K. Yasuda, A. Tsukazaki, K. S. Takahashi, N. Nagaosa, M. Kawasaki, Y. Tokura, Quantum Hall states stabilized in semi-magnetic bilayers of topological insulators. *Nat. Commun.* **6**, 8530 (2015).

31. J. Zhang, C. Z. Chang, Z. Zhang, J. Wen, X. Feng, K. Li, M. Liu, K. He, L. Wang, X. Chen, Q. K. Xue, X. Ma, Y. Wang, Band structure engineering in $(Bi_{1-x}Sb_x)_2Te_3$ ternary topological insulators. *Nat. Commun.* **2**, 574 (2011).

32. R. Yoshimi, A. Tsukazaki, Y. Kozuka, J. Falson, K. S. Takahashi, J. G. Checkelsky, N. Nagaosa, M. Kawasaki, Y. Tokura, Quantum Hall effect on top and bottom surface states of topological insulator $(Bi_{1-x}Sb_x)_2Te_3$ films. *Nat. Commun.* **6**, 6627 (2015).

33. V. M. Edelstein, Spin polarization of conduction electrons induced by electric current in two-dimensional asymmetric electron systems. *Solid State Commun.* **73**, 233-235 (1990).

34. Y. Zhang, K. He, C. Z. Chang, C. L. Song, L. L. Wang, X. Chen, J. F. Jia, Z. Fang, X. Dai, W. Y. Shan, S. Q. Shen, Q. Niu, X. L. Qi, S. C. Zhang, X. C. Ma, Q. K. Xue, Crossover of the three dimentional topological insulator $Bi_2Se_3$ to the two-dimentional limit. *Nat. Phys*. **6**, 584-588 (2010).

35. K. S. Lee, S. W. Lee, B. C. Min, K. J. Lee, Threshold current for switching of a perpendicular magnetic layer induced by spin Hall effect. *Appl. Phys. Lett.* **102**, 112410 (2013).

36. C. Zhang, M. Yamanouchi, H. Sato, S. Fukami, S. Ikeda, F. Matsukura, H. Ohno, Magnetization reversal induced by in-plane current in Ta/CoFeB/MgO structures with perpendicular magnetic easy axis. *J. Appl. Phys.* **115**, 17C714 (2014).

37. O. J. Lee, L. Q. Liu, C. F. Pai, Y. Li, H. W. Tseng, P. G. Gowtham, J. P. Park, D. C. Ralph, R. A. Buhrman, Central role of domain wall depinning for perpendicular magnetization switching driven by spin torque from the spin Hall effect. *Phys. Rev. B* **89**, 024418 (2014).

**Acknowledgements:** We thank S. Seki, F. Kagawa, H. Oike and M. Kawamura for fruitful discussions and experimental supports. **Funding:** This research was supported by the Japan Society for the Promotion of Science through the Funding Program for World-Leading Innovative R & D on Science and Technology (FIRST Program) on "Quantum Science on Strong Correlation" initiated by the Council for Science and Technology Policy and by JSPS Grant-in-Aid for Scientific Research(S) No. 24224009 and No. 24226002 and No. JP15H05853 from MEXT, Japan. This work was supported by CREST, JST.

**Author contributions:** Y. T. and M. K. conceived the project. K. Y. grew the topological insulator heterostructures, fabricated the device and performed measurement with the help of R. Y., A. T., K. K.,

K. S. T. and Y. O.. K. Y. analyzed the data and performed theoretical calculation. K. Y., A. T., Y. T. and M. K. jointly discussed the result and wrote the manuscript with contributions from all authors.

# FIGURE LEGENDS

**Fig. 1. Two origins of second harmonic Hall voltage.** (**A**) Schematic illustration of second harmonic Hall voltage $V_y^{2\omega}$ caused by spin-orbit torque mechanism in CBST/BST. External magnetic field $B$ and ac current $J^\omega$ are applied in the parallel direction. Magnetization $M$ is tilted by the effective field due to SOT $B_{SOT}$ caused by the spin accumulation $\sigma$. (**B**) Schematic illustration of $V_y^{2\omega}$ caused by $J^2$-proportional nonlinear conduction (asymmetric magnon scattering). Nonlinear conduction results in finite $V_y^{2\omega}$ at $M \parallel J$ configuration even if magnetization is not tilted from in-plane direction. (**C**) Magnetic field dependence of $V_y^{2\omega}$ at 2 K and 1 μA for the CBST/BST/InP sample. The inset shows the signal at low magnetic fields (< 2 T). (**D**) The same as **C** for the inverted sample, BST/CBST/InP. (**E**) Current magnitude $J$ dependence of $V_y^{2\omega}$ at 2 K and 1 T for the CBST/BST/InP sample. The dotted line represents $V_y^{2\omega} \propto J^2$.

**Fig. 2. Angular dependence of second harmonic voltage.** (**A** to **C**) Schematic illustrations of measurement configurations for angular dependence. (**D** to **F**) Angular dependence of second harmonic longitudinal voltage $V_x^{2\omega}$ and second harmonic transverse (Hall) voltage $V_y^{2\omega}$ for $zx$ plane (**D**), $zy$ plane (**E**) and $xy$ plane (**F**). The measurements were done at 2 K, 1 T and 1 μA. The green line $V_x^{2\omega, SOT}$ illustrates the calculated second harmonic voltage if $V_y^{2\omega}$ were assumed to originate purely from the SOT mechanism with magnetization oscillation. (**G**) Illustration of the origin of $V_x^{2\omega}$ under $J \perp M$ configuration. Red and pink arrows represent the scatterings by magnon emission and absorption processes at around the Fermi surface, respectively. (**H**) Illustration of the origin of $V_y^{2\omega}$ under $J \parallel M$ configuration.

**Fig. 3. Comparison between experiment and calculation of second harmonic Hall voltage caused by asymmetric magnon scattering.** (**A**) Observed temperature dependence of $V_y^{2\omega}$ under various magnetic fields at 1 µA under $J \parallel B$ configuration. (**B**) Numerical calculation results of temperature dependent $V_y^{2\omega}$ under various magnetic fields.

**Fig. 4. Magnetization switching via spin-orbit torque mechanism.** (**A**) Schematic illustration of current-pulse-induced magnetization switching. (**B**) Schematic of measurement procedure for current pulse $J_{pulse}$ induced magnetization switching. Hall resistance was measured under low current (~ 1 µA) at the green triangle every time after current pulse injection as shown in the gray bar. The current pulse $J_{pulse}$ is first applied in the positive direction, then in the negative direction, and then positive with gradually changing its magnitude. The pulse width is set to be ~ 1 ms. (**C**) Current pulse amplitude $J_{pulse}$ dependence of Hall resistance $R_{yx}$ under in plane magnetic field $B = \pm 0.02$ T at 2 K measured with the procedure shown in **B**. The corresponding current pulse density $j_{pulse}$ is shown on the upper abscissa. The normalized $M_z$, calculated as $R_{yx}/R_{AHE}$, is shown in the right scale, where $R_{AHE} = 2.3$ kΩ. This corresponds to $(r_{up} - r_{down})/(r_{up} + r_{down})$, where $r_{up}$ ($r_{down}$) is the fraction of up (down) domain. (**D**) Repeated magnetization switching under $B = \pm 0.02$ T and $J_{pulse} = \pm 1$ mA at 2 K. Prior to the first current pulse, magnetization is initialized to the multidomain state with zero net $M_z$ and the consecutive current pulses are applied.

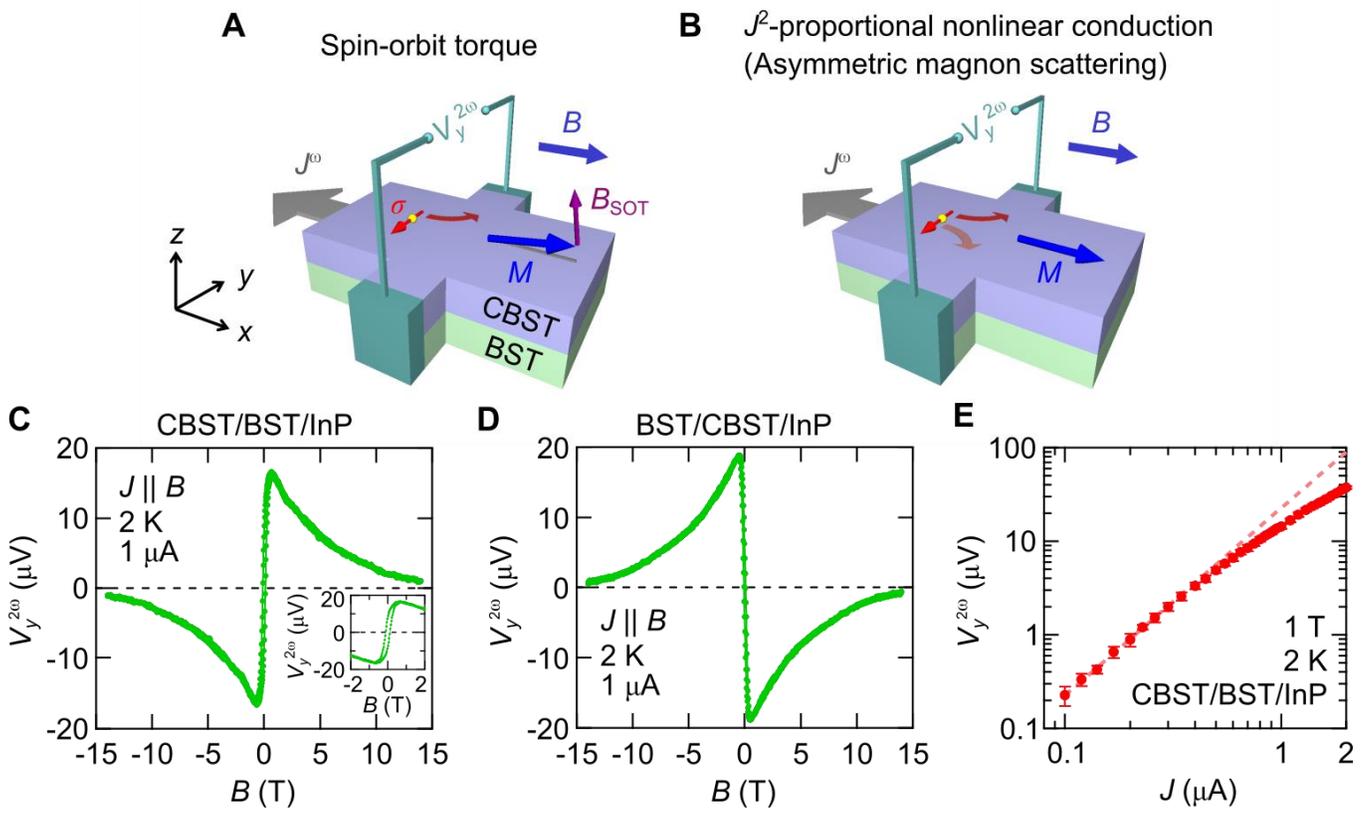

Fig. 1 K. Yasuda *et al*.,

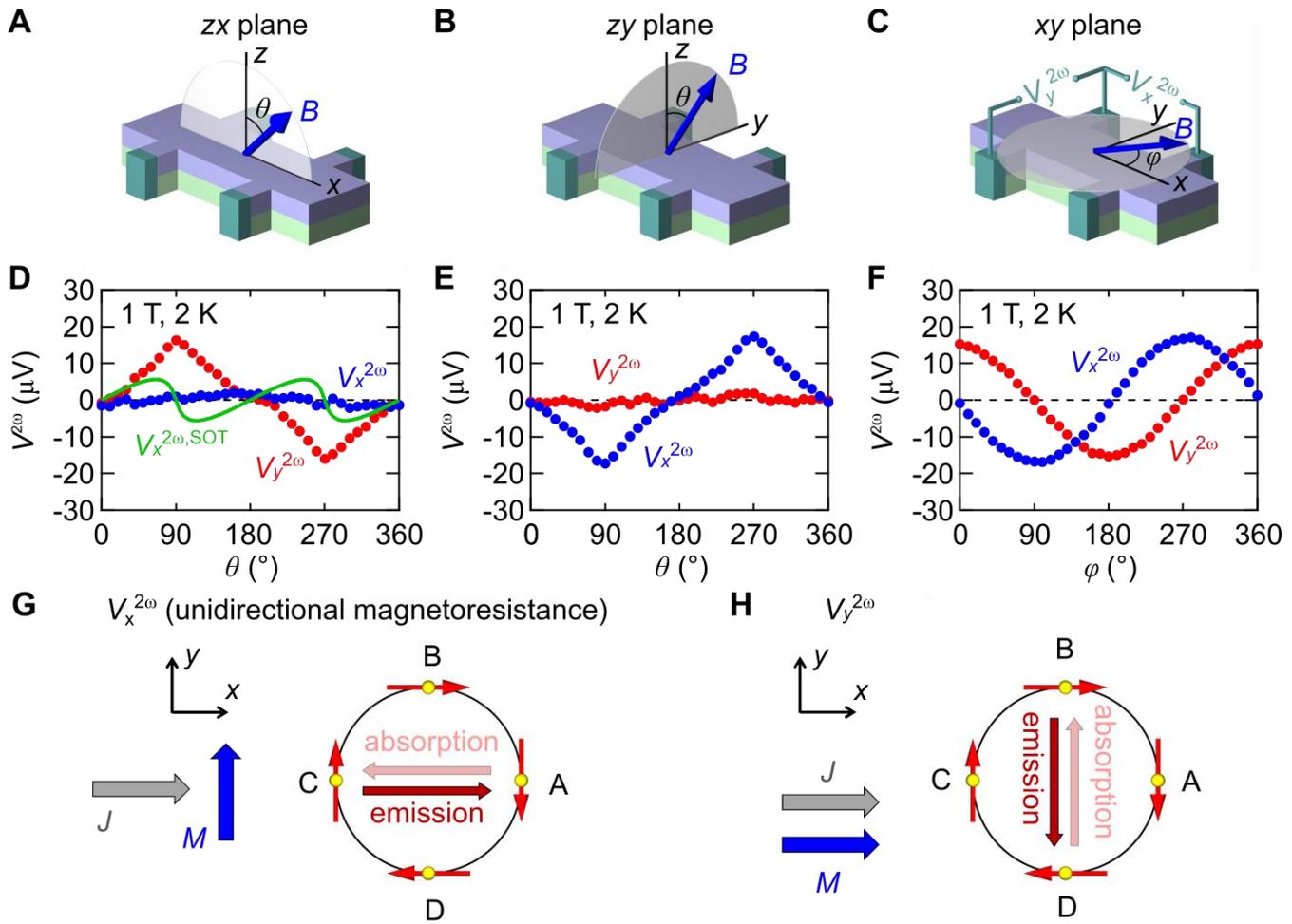

Fig. 2 K. Yasuda *et al*.,

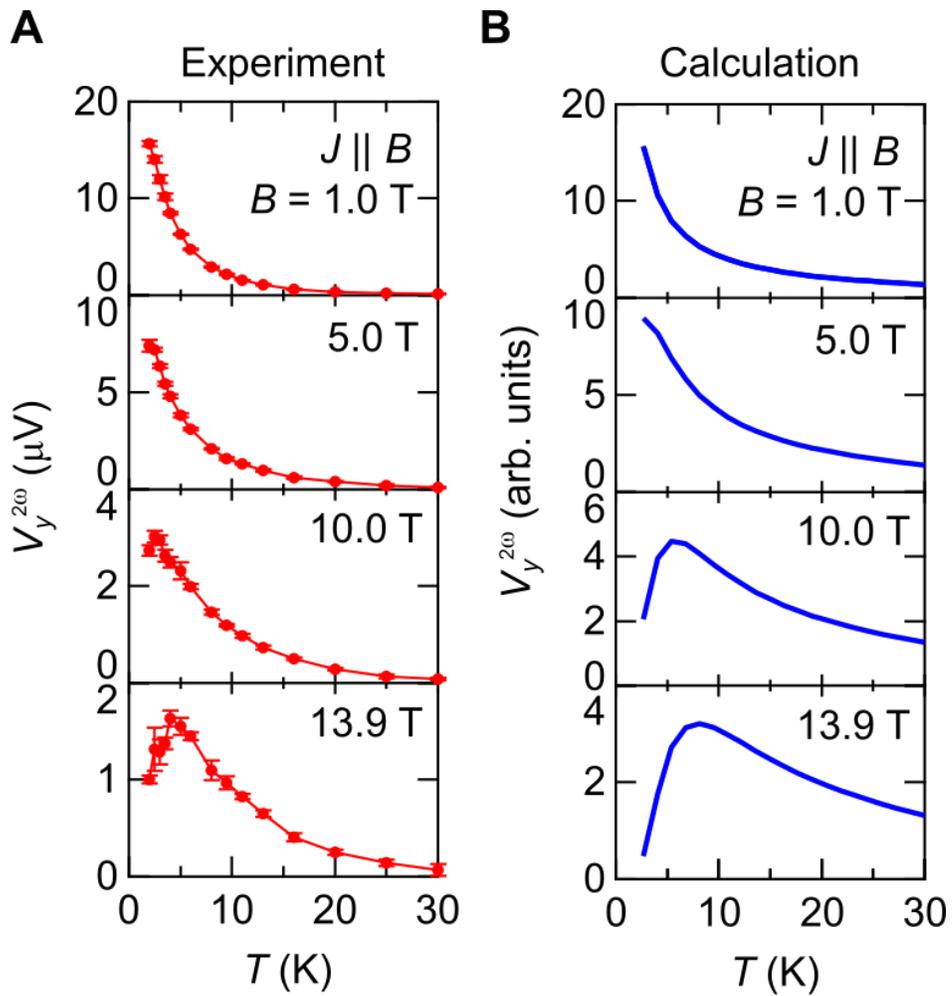

Fig. 3 K. Yasuda *et al.*,

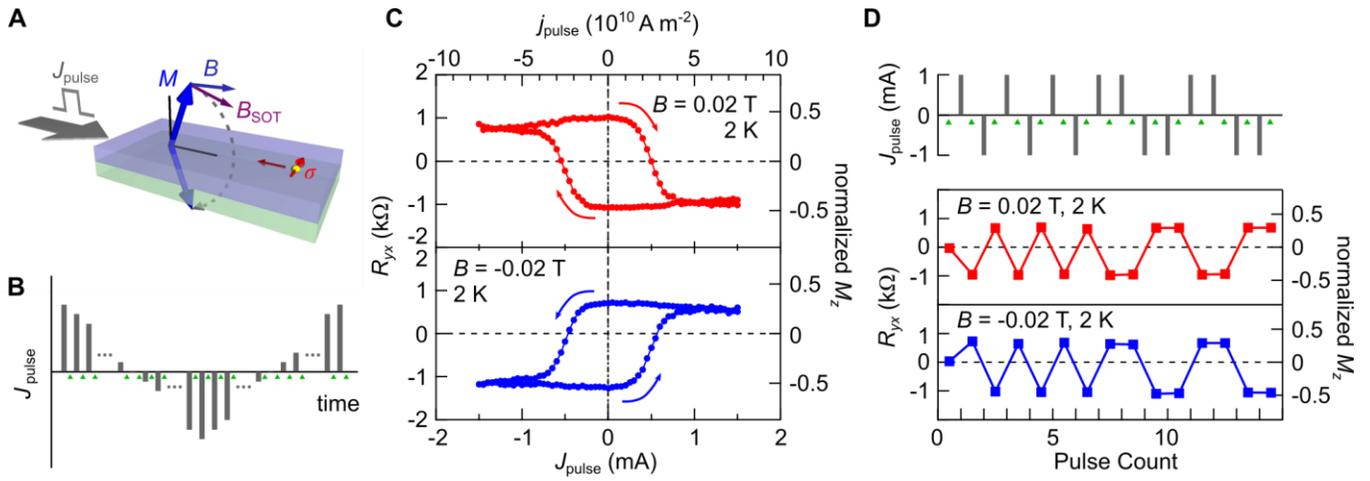

Fig. 4 K. Yasuda *et al.*,